\documentclass[aps, 10pt, amsmath, amssymb, prl, notitlepage, twocolumn]{revtex4-1}
\usepackage{mathtools}
\usepackage{young}
\usepackage{qcircuit}
\usepackage[size=small]{caption}
\usepackage{etoolbox}
\usepackage{appendix}
\usepackage{graphicx}   
\usepackage[utf8]{inputenc}
\usepackage[main=english, czech]{babel}
\usepackage[normalem]{ulem}
\usepackage{bm}
\usepackage{natbib}
\usepackage{amsthm}
\usepackage{braket}
\usepackage{proof}
\newtheorem{definition}{Definition}

\usepackage{url}
\usepackage{calc}
\usepackage{tikz-qtree}
\usetikzlibrary{trees, patterns}
\usetikzlibrary{positioning}
\usepackage{tikz,ifthen,calc}
\usepackage{subcaption}

\usepackage{complexity}
\usepackage{fancyref}
\usetikzlibrary{shapes,snakes}

\definecolor{ltgray}{rgb}{0.95,0.95,0.95}

\newcommand{\legenddiamond}[1][fill=black]{\tikz [x=1.2ex,y=1.85ex,line width=.1ex,line join=round, yshift=-0.285ex] \node[fill,diamond, inner sep = 2pt, fill=black] at (0,0) {};}%\draw  [#1] 
\newcommand{\legendcircle}[1][fill=black]{\tikz [x=1.2ex,y=1.85ex,line width=.1ex,line join=round, yshift=-0.285ex] \node[fill,circle,inner sep=2.5pt,fill=black] at (0,0) {};}%
\newcommand{\legendsmallcircle}[1][fill=black]{\tikz [x=1.2ex,y=1.85ex,line width=.1ex,line join=round, yshift=-0.285ex] \node[fill,circle,inner sep=1.5pt,fill=black] at (0,0) {};}%
\newcommand{\legendsquare}[1][fill=black]{\tikz [x=1.2ex,y=1.85ex,line width=.1ex,line join=round, yshift=-0.285ex] \node[fill,rectangle,inner sep=3pt,fill=black] at (0,0) {};}%

\pgfdeclarepatternformonly{soft horizontal lines}{\pgfpointorigin}{\pgfqpoint{100pt}{1pt}}{\pgfqpoint{100pt}{3pt}}%
{
  \pgfsetstrokeopacity{0.3}
  \pgfsetlinewidth{0.1pt}
  \pgfpathmoveto{\pgfqpoint{0pt}{0.5pt}}
  \pgfpathlineto{\pgfqpoint{100pt}{0.5pt}}
  \pgfusepath{stroke}
}

\usetikzlibrary{calc,intersections, fit, knots, hobby, positioning, patterns}
\usepackage{braids}

\usetikzlibrary{arrows}

\tikzset{
  treenode/.style = {align=center, inner sep=0pt, text centered,
    font=\fontsize{6}{22.4}\sffamily },
  arn_n/.style = {treenode, circle, white, draw=black,
    fill=black, text width=1em},
  arn_r/.style = {treenode, circle, white, draw=red, fill=red, 
    text width=1em, thick},
  arn_x/.style = {treenode, rectangle, draw=black,
    minimum width=0.5em, minimum height=0.5em}
}

\usepackage{hyperref}
\hypersetup{
    colorlinks=true,
    linkcolor=blue,  
    citecolor=blue,
    urlcolor=blue,
}
 
\begin{document}
\title{Quantum Schur Sampling Circuits can be Strongly Simulated}

\begin{abstract}
Permutational Quantum Computing (PQC) [\emph{Quantum~Info.~Comput.}, \textbf{10}, 470--497, (2010)] 
is a natural quantum computational model conjectured to capture non-classical aspects of quantum computation. An argument backing this conjecture was the observation that there was no efficient classical algorithm for estimation of matrix elements of the $S_n$ irreducible representation matrices in the Young's orthogonal form, which correspond to transition amplitudes of a broad class of PQC circuits. This problem can be solved with a PQC machine in polynomial time, but no efficient classical algorithm for the problem was previously known. Here we give  a classical algorithm that efficiently approximates the transition amplitudes up to polynomial additive precision and hence solves this problem. We further extend our discussion to show that transition amplitudes of a broader class of quantum circuits -- the Quantum Schur Sampling circuits --  can be also efficiently estimated classically. 
\end{abstract}

\author{\foreignlanguage{czech}{Vojtěch Havlíček}}
\email{vojtech.havlicek@keble.ox.ac.uk}
\affiliation{Quantum Group, Department of Computer Science, University of Oxford, Wolfson Building, Parks Road, Oxford
OX1 3QD, UK}

\author{Sergii Strelchuk}
\affiliation{Department of Applied Mathematics and Theoretical Physics, University of Cambridge, Wilberforce Road,  Cambridge, CB2 3HU, UK} 

\maketitle

The effort for building quantum computers is driven by a belief that they can perform computational tasks beyond capabilities of classical computers. Despite a plethora of evidence for this conjecture, it remains unclear what kinds of quantum algorithms would lead to such computational advantage. With the development of nascent proof-of-principle quantum devices, there is an increased demand for simple and useful quantum protocols which could clearly demonstrate their supra-classical capabilities. This motivated various attempts to delineate computational power of the near-term devices in the regime where quantum error correction is not readily available~\cite{Bremner11, Bremner16, Harrow17}. One such direction is the study of restricted quantum computation that aims to identify sources of quantum advantage within computational models that are likely to be less powerful than universal quantum computation. 

We study Permutational Quantum Computing (PQC) introduced by Marzuoli and Rasetti and further developed by Jordan ~\cite{Marzuoli05, Jordan09} -- a natural quantum computational model that is conjectured to capture non-classical aspects of quantum computation. The backing for this conjecture came from several computational problems which were solvable on a PQC machine in polynomial time but the only known classical algorithms required exponential runtime. We provide an efficient classical probabilistic algorithm for polynomially small approximation of transition amplitudes of a class of PQC computations that allows us to resolve one of the problems used to back up the conjecture of the model supra-classicality. While this does not fully resolve the computational power of the model, it should be seen as a step towards isolating features of the model that could be responsible for its quantum computational advantage. Furthermore, we extend this result to a larger class of important quantum circuits -- the so-called Quantum Schur Sampling circuits -- which get their name from bearing structural similarities to circuits that perform Quantum Fourier Sampling \cite{Fefferman15}. 

Following the applications of PQC outlined by Jordan \cite{Jordan09}, the proposed algorithm can be used not only to ascertain the power of the computational model but also to potentially address problems of practical relevance: it can be used to estimate non-trivial elements of irreducible representation matrices of the symmetric group on $n$ elements in the Young-Yamanouchi basis \cite{Pauncz67}, which could find applications in quantum chemistry and various quantum information protocols. 

PQC took inspiration in Topological Quantum Computing; a proposal for quantum computation by anyonic braiding and measurement in which results of the computation are determined solely by the topology of anyon trajectories. This approach relies on the existence of non-abelian anyons which have not been experimentally confirmed and its viability thus remains unknown. Permutational Quantum Computing can be then seen as an attempt to mimic some of this structure in a qubit-based model, in which angular momentum (spin) eigenstates roughly play the role of anyons and the computation proceeds by qubit interchange and spin measurements.  The analogy between the two models is however very loose, as PQC completely disregards topology by forgetting about anyonic trajectories and only accounts for particle permutations; hence its name. 

PQC takes input in a basis defined by a set of spin measurements on qubit subsets. This basis, similarly to Fourier basis, is defined by outcomes of nonlocal measurements and is therefore not a tensor product basis. 
Nevertheless, there exists an efficient quantum algorithm to prepare the basis states. It is also possible that a suitable sequence of spin measurements could be implemented directly experimentally \cite{Jordan09}.

We now describe the basis. Consider $n$ two-level quantum systems or qubits.  With a convention that $\hbar = 1$, a spin of $k$-th qubit is formally defined by a triple:
\begin{align*}
\vec{S}_k &= \frac{1}{2} \left( X_k, \, Y_k, \,Z_k \right),
\end{align*}
where $X_k$ denotes a Pauli $X$ operator applied to $k$-th qubit. The total spin of a subset of qubits  $A$  is then given by: \[ S_A^2 = \left( \sum_{k \in A} \vec{S}_k \right)^2\,.\] Let $Z_A :=  \frac{1}{2} \sum_{k \in A} Z_k $ 
denote the $z$-spin operator on a qubit subset $A$ and let $Z$ be such operator applied to all of the $n$ qubits.  The operators $Z_A$ and $S^2_A$ commute and share eigenspace labeled by quantum numbers $j_A$~and~$m_A$.

Here $j_A$ is the total spin of qubits in $A$ and $m_A$ is the multiplicity label taking values in integer steps between $ -j_A $ and $ j_A$. The $j$-quantum numbers are either integer or half-integer and combine according to the angular momentum addition rules~\cite{Woit17}: 
\begin{align*}
 |j_A - j_B| \leq j_{A \cup B} &\leq j_A + j_B, & 
 j_{A \cup B} + j_A + j_B & \in \mathbb{Z}.
\end{align*} 

 The operators $S_A^2$ and $S_B^2$ can be shown to commute if and only if $A$ and $B$ are disjoint or one is subset of the other. An orthonormal basis can be then specified by listing outcomes of a complete set of commuting spin measurements on qubit subsets. Specifically, a basis can be defined by coupling one qubit at a time - that is by joint eigenstates of operators $\, S_{\lbrace 01 \rbrace}^2, \, S_{\lbrace 012 \rbrace}^2, \,S_{\lbrace 0123 \rbrace}^2, \, \ldots S^2$ along with the $Z$ operator and will be referred to as \textit{sequentially coupled basis}. An example of a sequentially coupled basis state on three qubits would be:
{\small \[ \Ket{J = \frac{1}{2}, M = \frac{1}{2}, j_{01} = 1}_\textsf{Sch} = \sqrt{\frac{2}{3}} \ket{001} -  \frac{ \ket{010} + \ket{100} }{\sqrt{6}}. \]}

Since the basis change does not alter the Hilbert space dimension the basis states can be assigned a bitstring $x \in \lbrace 0, 1 \rbrace^n$ and we will label the sequentially coupled basis state corresponding to a bitstring $x$ by $\ket{x}_{\textsf{Sch}} \equiv \ket{J,M, \ldots}_{\textsf{Sch}}$ where appropriate. 
Note that the $j$ values label spaces arising in the decomposition of the $n$-qubit Hilbert space into a direct sum of $\textsf{SU}(2)$ irreducible representations. Representation-theoretically, the sequentially coupled basis is the Young-Yamanouchi basis~\cite{Pauncz67, Jordan09}. %It remains a The proposed classical simulation algorithm however works for arbitrary basis specified by a complete set of the spin subset measurements {\color{blue} [Discuss generalization to arbitrary basis]}.

The unitary transformation $U_{\textsf{Sch}}$ between the computational and the sequentially coupled basis is called the Quantum Schur Transformation. Similarly to the Quantum Fourier Transform, it is a global transformation from the computational basis to the basis labeled by the spin eigenvalues~\cite{Bacon06}. Recall that states in two spin eigenspaces $j_L$ and $j_R$ couple to a spin eigenstate $\ket{j,m}$ as:
\[ \ket{j,m} := \sum_{m_L, m_R}  C_{j_L, m_L; j_R, m_R}^{j,m} \ket{j_L, m_L} \otimes \ket{j_R, m_R} \,, \] 
where $C^{j,m}_{j_L,m_L; j_R, m_R}$ are the Clebsch-Gordan (CG) coefficients. Note also that one can identify the computational basis states with a tensor product of spin eigenstates by:
\begin{align} \Ket{\frac{1}{2}, m =-\frac{1}{2}} &\equiv \ket{0}, & \Ket{\frac{1}{2}, m =\frac{1}{2}} &\equiv \ket{1}. \label{corresp} \end{align}
Since the spin of a single qubit is $j = \frac{1}{2}$, we will drop the $j$ label for qubits where appropriate to reduce notational clutter. The Schur transform on $n$ qubits $\ket{m_0}\ldots\ket{m_{n-1}}$ in the sequentially coupled basis starts by coupling $\ket{m_0}$ with $\ket{m_1}$ into $\ket{j_{01}, m_{01}}$, and proceeds by coupling $\ket{m_2}$ with $\ket{j_{01}, m_{01}}$ into $\ket{j_{012}, j_{01}, m_{012}}$. This state still carries the label $j_{01}$ since the total angular momentum operators $S_{\lbrace 012 \rbrace}$ and $S_{\lbrace 01 \rbrace}$ commute. In contrast, the label $m_{01}$ is discarded, because $Z_{\lbrace 01 \rbrace}$ no longer commutes with $S_{\lbrace 012 \rbrace}$. This process is continued until the whole register is coupled into a sequentially coupled basis state $\ket{J, M, \ldots j_{012}, j_{01}}_\textsf{Sch}$. The transformation can be efficiently implemented using the algorithms of Bacon, Harrow and Chuang and Kirby and Strauch (see~\cite{Childs06, Bacon06, Bacon06b, Kirby17} for details).

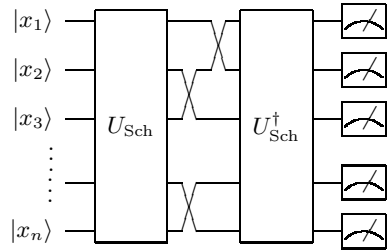
\begin{figure}[t]
\[
\Qcircuit @C=0.6em @R=0.6em {
   & \lstick{\ket{x_1}}     & \qw & \multigate{5}{U_{\text{Sch}}} & \qw & \qw          & \qw & \link{1}{-1} &\multigate{5}{U^\dag_{\text{Sch}}} & \qw & \meter \\
   & \lstick{\ket{x_2}}     & \qw & \ghost{U_{\text{Sch}}}        & \qw & \link{1}{-1}  & \qw &  \link{-1}{-1}  &\ghost{U^\dag_{\text{Sch}}}        & \qw &\meter \\
   & \lstick{\ket{x_3}}     & \qw & \ghost{U_{\text{Sch}}}        & \qw & \link{-1}{-1} &  \qw & \qw &\ghost{U^\dag_{\text{Sch}}}        & \qw &\meter \\
  \vdots  &                     & \      &                                            &       &                    &  & &                                                   &  & \\
   & \lstick{\vdots} & \qw & \ghost{U_{\text{Sch}}}        & \qw & \link{1}{-1}  & \qw & \qw & \ghost{U^\dag_{\text{Sch}}}        & \qw &\meter \\
   & \lstick{\ket{x_n}}     & \qw & \ghost{U_{\text{Sch}}}        & \qw & \link{-1}{-1}  & \qw & \qw &\ghost{U^\dag_{\text{Sch}}}        & \qw &\meter \\
}
\]
\caption{Permutational Quantum Computing. The `program' $\pi$ consists of a set of permutations executed between $U_{\textsf{Sch}}$ and $U^\dag_{\textsf{Sch}}$.}
\end{figure}

We now define the computational model.
Given a permutation $\pi$ on $n$ qubit labels and a classical input $x \in \lbrace 0,1 \rbrace^n$: 
\begin{enumerate}
\item Prepare a sequentially coupled basis state $U_{\textsf{Sch}} \ket{x} = \ket{x}_{\textsf{Sch}} \equiv \ket{ j_{01}, \ldots, J, M}_\textsf{Sch}$ by applying the Quantum Schur Transformation $U_{\textsf{Sch}}$ to a computational basis state $\ket{x}$. 

\item Execute the `program'  $\pi$, which permutes the qubits (or equivalently, their labels). That is: 
\begin{align*} \ket{j_{01}, \ldots, J, M}_\textsf{Sch} \mapsto \ket{j_{\pi(0)\pi(1)}, \ldots, J, M}_\textsf{Sch}. \end{align*}
This transformation can be also regarded as a sequence of computational basis \textsf{SWAP} gates that has depth $O(n^2)$ by bubblesort algorithm \cite{Jordan09}. The operation acts as a unitary $U_\pi$ on the input state.
\item Apply the inverse Quantum Schur Transform  $U_{\textsf{Sch}}^\dag$ to get a superposition over output states $\ket{y}$.

\item Measure the $y \equiv J,M, j' \ldots$ labels.
 \end{enumerate}

After executing steps 1-4, the algorithm samples from a probability distribution $ p_\pi(y|x) $ over the output $y$.

We now show that the transition amplitudes of Permutational Quantum Computing circuits that use the sequentially coupled basis can be efficiently approximated classically. It is known that the PQC transition amplitudes in this regime correspond to the matrix elements of the irrep. matrices of the symmetric group in Young's orthogonal form, so resolving this problem provides a classical algorithm for task that has been conjectured hard in \cite{Jordan09} (See \cite{Jordan09} for additional details of the correspondence). Our proof strategy follows work of van den Nest by showing that the output state is {\it computationally tractable} and invoking a theorem about classical simulation of overlaps of such states~\cite{VanDenNest09}. 

\begin{definition}[Computational Tractability~\cite{VanDenNest09}]\label{def:tractability}
An $n$-qubit state $\ket{\psi}$ is computationally tractable if it is possible to classically efficiently  sample from $p = \lbrace |\braket{y | \psi} |^2: y \in \lbrace 0,1\rbrace^n \rbrace$ and the overlaps $\braket{y | \psi}$ can be computed classically up to $m$ significant bits in time $poly(n,m)$ for any computational basis state $\ket{y}$.  
\end{definition}

We first prove that the state $\braket{y | J,M, \ldots}_{\textsf{Sch}} $ can be efficiently computed classically to the required precision. To simplify the presentation, we work with either three or four qubits -- the method can be straightforwardly generalized to an arbitrary number of qubits. By angular momentum conservation, the CG coefficients satisfy: 
\begin{align} C^{J,M}_{j_L, m_L; j_R, m_R} &= \delta^M_{m_L + m_R} C^{J,M}_{j_L, m_L; j_R, m_R}. \label{conservation} \end{align}
It follows that a $\ket{J,M, j_{01}}_\textsf{Sch}$ state on three qubits can be written as:
\begin{align*} &\ket{J,M,j_{01}}_\textsf{Sch} :=  \\  
&\sum_{m_{01}, m_0, m_1, m_2} C^{J,M}_{j_{01}, m_{01}; m_2} C^{j_{01},m_{01}}_{m_0; m_1} \ket{m_0 m_1 m_2}\\ &=
\sum_{m_0, m_1, m_2} C^{J,M}_{j_{01}, m_0 + m_1; m_2} C^{j_{01},m_0+m_1}_{m_0; m_1} \ket{m_0 m_1 m_2},\end{align*}
where Eq.~\ref{conservation} was used to cancel the summation over $m_{01}$.
Label $M_l := \sum_{k=0}^l m_k$ for notational convenience and recall that a computational basis state $\ket{y}$ for $y \in \lbrace 0, 1 \rbrace^4$ is equivalent to $\ket{m_0m_1m_2m_3}$ by the correspondence in Eq.~\ref{corresp}. Given a sequentially coupled basis state $\ket{J,M,j_{012}, j_{01}}_{\textsf{Sch}}$ on four qubits, the overlap $\braket{y | J, M, j_{012}, j_{01}}_{\textsf{Sch}}$ is hence given by:
\begin{align*}
&\Braket{y |J, M,j_{012},j_{01}}_{\textsf{Sch}} =\\
&\Braket{m_0m_1m_2m_3 |J, M,j_{012},j_{01}}_{\textsf{Sch}} = \\
& C^{J, M}_{j_{012}, M_2; m_3}\;  C^{j_{012}, M_2}_{j_{01}, M_1; m_2} \; C^{j_{01}, M_1}_{m_0; m_1}. \end{align*}
The summation over the intermediate $m$ numbers again vanishes due to angular momentum conservation of Eq.~\ref{conservation}. The state overlap $\braket{y|J,M, \ldots }_{\textsf{Sch}}$ on $n$ qubits can be hence evaluated if it is possible to efficiently compute a product of at most $(n-1)$ \textsf{SU}(2) CG coefficients. These can be computed using:
\begin{align*}
C^{J,M}_{j_1, m_1; j_2, m_2} &= (-1)^{M + j_1 - j_2}\sqrt{2J + 1} 
\left(\begin{array}{ccc} j_1 & j_2 & J \\ m_1 & m_2 & M \end{array} \right),
\end{align*}
where the six-symbol array denotes the Wigner $3j$ symbol.
Since the total angular momenta ($j$-numbers) are all upper-bounded by $\frac{n}{2}$ by the angular momentum addition rules, 
the Wigner $3j$ symbols can be computed in polynomial time to exponential precision using the Racah formula~\cite{racah1942theory}. This allows for efficient evaluation of the Clebsch-Gordan coefficients
up to the $m$ significant bits in $poly(n,m)$ time (see the Supplemental Material for additional details). Thus, a product of $(n-1)$ CG coefficients can be evaluated efficiently to the necessary precision.

We now prove the second computational tractability condition, that is that the probability distributions arising from computational basis measurements on the sequentially coupled states can be sampled classically in polynomial time. We first show that it is possible to efficiently compute the `telescoping' marginals of the output probability distribution $p(y) := |\braket{y | J,M \ldots}_\textsf{Sch}|^2$, which are defined by: 
\[ p(y_k \ldots y_{n-1}) = \sum_{y_0 \ldots y_{k-1}} p(y_0 \ldots y_{n-1}).  \]
This can be achieved by successive application of:
\begin{align*} &p(y_k \ldots y_{n-1}) = \sum_{y_0 \ldots y_{k-1}} p(y_0 \ldots y_{n-1}) \\ &= \sum_{m_0 \ldots m_{k-1}} \left| C^{J, M}_{j_{n-2 \ldots 0}, M_{n-2}; m_{n-1}} C^{j_{n-2 \ldots 0}, M_{n-2}}_{j_{n-3 \ldots 0}, M_{n-3}; m_{n-2}} \ldots \right|^2 \\
&=  \left| C^{J, M}_{j_{n-2 \ldots 0}, (M - m_{n-1}); m_{n-1}} \right|^2 \sum_{y_{0} \ldots y_{k-1}} p(y_0 \ldots y_{n-2}),
\end{align*}
where we again used the correspondence of Eq.~\ref{corresp} and the angular momentum conservation of Eq.~\ref{conservation} was used to proceed from the second to the third line - recall that $M$ is fixed as the basis states are eigenstates of $Z$ with this eigenvalue and that $m_{n-1}$ and hence $M-m_{n-1}$ is also fixed, since $m_{n-1}$ is known. In contrast, the $M_k$ were all \textit{a priori} summed over. The substitution hence allows to pull a CG factor in front of the above summation. 
Repeating this $(n-k)$ times and noting that by normalization: 
\[ \sum_{y_0 \ldots y_{k-1}} p(y_0 \ldots y_{k-1}) = 1, \]
gives the desired marginal as a factor of at most $(n-1)$ CG coefficients. Hence the telescoping marginals are also polytime classically computable to the desired precision.

The output distribution on $n$ qubits can be then efficiently sampled from by using the chain rule for probability distributions: 
{\small
\begin{align*}
p(y_0 \ldots y_{n-1}) &=  p(y_0|y_1 \ldots y_{n-1}) \ldots p(y_{n-2}|y_{n-1}) p(y_{n-1}),
\end{align*}}
which implies that the output probability distribution can be computed by successively drawing individual bits. First, sample $y_{n-1}$, which can be done by flipping a coin with bias given by the telescoping marginal $p(y_{n-1})$.  Fix $y_{n-1}$ to the result, and generate a sample $y_{n-2}$ by flipping a coin with bias $p( y_{n-2} | y_{n-1} )$. Proceed the same way up to $y_0$. The resulting bitstring $y= y_0 y_{1} \ldots y_{n-1}$ of the results is then drawn from $p(y) = p(y_0 y_{1} \ldots y_{n-1})$ by the chain rule, as desired. One can hence sample the output in polynomial time classically. 

We can now prove that an output probability distribution of the Permutational Quantum Computing in the bases considered here can be approximated in polynomial time classically. Consider a Permutational Quantum Computing circuit of the form:
\begin{align*}
p_\pi (y|x) &= \left| \bra{y_0 \ldots y_n} U^\dag_{\textsf{Sch}} U_\pi \ket{J,M, j_{012}, j_{01}}_\textsf{Sch} \right|^2  \\
&= \left| \braket{J,M, j_{01}, j_{012} | J,M, j_{\pi{(0)}\pi{(1)}\pi(2)}, j_{\pi(0) \pi(1)}}_\textsf{Sch} \right|^2 \\
&= \left| \braket{\phi | \psi} \right|^2,
\end{align*}
where $U_\pi$ implements the permutation $\pi$ on qubit indices.
From the above, $\ket{\phi}$ and $\ket{\psi}$ are computationally tractable (this follows directly from Definition~\ref{def:tractability} by substituting $\ket{y}$ with $\pi\ket{y}$) and yield probability distributions:
\begin{align*}
p &:= |\braket{y|\phi}|^2, & p_{\pi} &:= |\braket{y|\psi}|^2 \,.
\end{align*}
As in Theorem 3 of~\cite{VanDenNest09}, consider an indicator function:
\[ \Upsilon(x) = \begin{cases} 1 \text{ if } p(y) > p_\pi(y), \\  0 \text{ if } p(y) \leq p_\pi(y), \end{cases}  \] and define:
{\small
\begin{align*}
F(x) &:= \frac{\braket{\phi | y} \braket{y | \psi}}{p(y)} \Upsilon(y),   & G(y) &:= \frac{\braket{\phi | y} \braket{y | \psi}}{p_\pi (y)} (1- \Upsilon(y)  )\,.
\end{align*}}
Both of these functions can be evaluated efficiently to the required precision, since $\ket{\psi}$ and $\ket{\phi}$ are both computationaly tractable. Then: 
\begin{align}
\braket{\phi | \psi} 
&= \sum_y p(y) F(y) + \sum_y p_\pi(y) G(y)\,.
\label{Eq:overlap}
\end{align}
The overlaps $\braket{\phi | \psi}$ are all real-valued as they are products of the CG coefficients, which are real. It follows there is a polynomially-precise additive approximation algorithm to the state overlaps by the Chernoff-Hoeffding bound:

\begin{definition}[Chernoff-Hoeffding bound~\cite{Schwarz13}]
Let $K_0, K_1, \ldots K_T$ be i.i.d. samples from a real random variable $K$, such that $|K_i| \leq 1$. Let $\langle K \rangle$ be the expectation value of $K$. Then for $T = \frac{2}{\epsilon^2} \log \left( \frac{2}{\delta} \right)$:
\begin{align*}
Pr \left( \left| \frac{1}{T} \sum_k^T K_k - \langle K \rangle \right| \leq \epsilon\right) &\geq 1 - \delta \,.
\end{align*}
\end{definition}

This implies that $O \left(\frac{1}{\epsilon^2} \log(\frac{1}{\delta})\right)$ samples suffice to give a polynomially precise approximation to both expectation values $\langle F \rangle := \sum_y p(y) F(y)$  and $\langle G \rangle = \sum_y p_\pi (y) G(y)$ with the accuracy $\epsilon$ and failure rate $\delta$. Setting $\epsilon = \frac{1}{poly(n)}$ hence gives an efficient simulation algorithm with exponentially small failure rate $\delta$. 

The analysis can be further extended to the case when the quantum circuit between the two Schur transformations is a composition of the permutation $\pi$ along with $Z$-diagonal circuit $\Lambda$ with classically efficiently computable elements. 
The output probabilities in the computational basis then remain unchanged as the elements of $\Lambda$ are pure phases: 
\begin{align*} p(y) &:= |\bra{y} \Lambda \ket{J,M, \ldots}_\textsf{Sch}|^2 \\ &= |\Lambda(y)|^2 |\braket{y|J,M, \ldots}_\textsf{Sch}|^2  
\\& = |\braket{y|J,M, \ldots}_\textsf{Sch}|^2. \end{align*}
This implies that the distributions from which $F, G$ are drawn in
Eq.~\ref{Eq:overlap} remain unchanged - although $F,\, G$ become complex-valued. The only difference this makes for the presented algorithm is that in definition 2, both real and complex part of the amplitude have to be estimated at the same time.

Such quantum circuits are structurally similar to Quantum Fourier Sampling circuits - except the Fourier transform is now replaced by the Schur transform. 

This observation hints at the importance of considering the presented classical algorithm when devising new quantum algorithms or arguments for quantum computational advantage.  

In this letter, we studied the computational power of the Permutational Quantum Computing and found an efficient classical simulation for a wide class of problems which are efficiently solvable within this model. This allowed us to find a solution to a problem of approximating matrix elements of the $S_n$ irreps in the Young's orthogonal form, for which there was previously no classical efficient solution~\cite{Jordan09}.

We now discuss some open questions. While we presented method for computing transition amplitudes for a broad class of PQC circuits, finding a method for sampling their output distribution remains open. 

As we studied computational power of PQC in a regime where the input/output states are encoded in the sequentially-coupled basis, it also remains open if our construction generalizes to arbitrary bases defined by complete commuting sets of spin subset measurements as the telescopic marginal method does not generalize simply to this case. 

It is also an open question to fully characterize admissible quantum gates that extend the notion of the `program' $\pi$ in step 2 beyond permutations or  $Z$-diagonal gates, such that the classical tractability of the resulting quantum state remains preserved. It would be additionally interesting to further explore connections to the proposal of a computational model based on permuting distinguishable quantum particles in superposition by Aaronson et al.~\cite{ABKM16}, as PQC is one of the limiting cases of this model. Lastly, we note that most of the ideas behind the PQC model originated from the spin networks~\cite{Penrose71, Marzuoli05}. It would be interesting to study further interplay of our result with this framework.

\begin{acknowledgments}
We acknowledge discussions with S.P.~Jordan, S.~Mehraban and G.~de Felice. We are grateful to an anonymous referees for pointing out technical slip-ups in the previous version of the paper. S.~Strelchuk is supported by a Leverhulme Trust Early Career Fellowship. V.~Havlicek is supported by Clarendon and Keble de Breyne scholarships.
\end{acknowledgments}

\bibliographystyle{unsrt}
\bibliography{main}

\newpage
\begin{widetext}
\section{Suplemental Material}
\paragraph{Evaluation of the CG coefficients.}
Here we describe how to efficiently evaluate the CG coefficients classically to the exponential precision. The CG coefficients are related to the Wigner $3j$ symbol as follows:
\begin{align}\label{cjfull}
C^{J,M}_{j_1, m_1; j_2, m_2} &= (-1)^{M + j_1 - j_2}\sqrt{2J + 1}
\left(\begin{array}{ccc} j_1 & j_2 & J \\ m_1 & m_2 & M \end{array} \right).
\end{align}
The Wigner $3j$ symbol can be computed using the Racah formula~\cite{racah1942theory}:
\begin{align}
\label{racah}
\left(\begin{array}{ccc} a & b & c \\ d & e & f \end{array} \right) = \sum_t (-1)^{t+a-b-f} \frac{\sqrt{\Delta{(abc)} (a+d)!(b+e)!(c+f)! (a-d)!(b-e)!(c-f)!}}{t!(c-b+t+d)!(c-a+t-e)!(a+b-c-t)!(a-d-t)!(b-t-e)!},
\end{align}
where $\Delta(abc)$ is the triangle coefficient given by:
\begin{align*}
\Delta(abc) &= \frac{(a+b-c)! (a-b+c)!(-a+b+c)!}{(a+b+c+1)!},
\end{align*}
and the summation $\sum_t$ runs over all  $t$ for which the factorials in the expression are well-defined. It is known that there are $\nu + 1$ such terms, where~\cite{Wigner3j}:
\begin{align*}
\nu &= \min \left \lbrace a \pm d, b \pm e, c \pm f, a+b-c, b+c-a, c+a-b  \right\rbrace\,. 
\end{align*}
In the context considered here $a,b \ldots f$ are total angular momenta of $n$ qubits - so it always holds that $a,b \ldots f \leq \frac{n}{2}$. Hence, the summation in Eq.~\ref{racah} runs over $poly(n)$ terms. Each of these terms only requires evaluation of factorials, division and square root - each of these operations can be performed efficiently up to $m$ bits in $poly(m)$ time. Evaluation of the Clebsch-Gordan coefficient up to $m$ significant bits will then take $poly(n,m)$ time, as required by the simulation algorithm.

\medskip

\end{widetext}

\end{document}